\documentclass[printer]{aa}  
\usepackage{graphics,epsf,epsfig,graphicx}  
\newcommand{\be}{\begin{equation}}
\newcommand{\ee}{\end{equation}}

\begin{document}

  \title{ Lyman-alpha radiative transfer during the Epoch of Reionization: contribution to 21-cm signal fluctuations }

\author{B. Semelin \inst{1}, F. Combes \inst{1} \and S. Baek \inst{1}}
\offprints{B. Semelin, \email{ benoit.semelin@obspm.fr}}
\institute{
LERMA, Observatoire de Paris, UPMC, CNRS, 61 Av. de l'Observatoire,
F-75014, Paris, France
}
\date{Received ; accepted }
\authorrunning{Semelin, Combes \& Baek}
\titlerunning{Lyman-alpha radiative transfer during the EoR}

\abstract{During the epoch of reionization, Ly-$\alpha$ photons emitted by the first stars
can couple the neutral hydrogen spin temperature to the kinetic gas temperature, providing the
opportunity to observe the gas in emission or absorption in the 21-cm line. Given the bright foregrounds,
it is of prime importance to determine precisely the fluctuations signature of the signal, to be able
to extract it by its correlation power.
LICORICE is a Monte-Carlo radiative transfer code, coupled to the dynamics via an adaptative Tree-SPH code.
We present here the Ly-$\alpha$ part of the implementation, and validate it through three classical tests.
 Contrary to previous works, we do not assume that $P\alpha$, the number of scatterings of
 Ly-$\alpha$ photons per atom per second, is proportional to the Ly-$\alpha$ background flux,
but take into account the scatterings in the Ly-$\alpha$ line wings. The latter have the effect
to steepen the radial profile of  $P\alpha$ around each source, and re-inforce the contrast
of the fluctuations. In the particular geometry of cosmic filaments of baryonic matter,
 $Ly-\alpha$ photons are scattered out of the filament, and the large scale structure of
 $P\alpha$ is significantly anisotropic. This could have strong implications for the possible
detection of the 21-cm signal.
}  
  
\maketitle  
  
\section{Introduction}  

The Epoch of Reionization (EoR) extends from the time when the first sources form as a result of
the nonlinear growth of primordial density fluctuations in a fully neutral universe ($z \sim 20$), to the 
moment when the intergalactic medium is fully reionized under the effect of UV radiations emitted
by the 
sources ($z\sim 6$). This simple picture, however, hides a number of uncertainties. At this time, there are 
only two observational constraints on the EoR. The first comes from the 
Gunn-Peterson effect in the absorption spectrum of high redshift quasars. Indeed the 
transmitted flux drops sharply as a small neutral fraction appears toward high redshifts:
it implies an ionization fraction $x_{\mathrm{HI}}< 10^{-4}$ at $z < 5.5$ (Fan {\sl et al.} 2006). 
The second constraint is set by the Thomson scattering of CMB photons  by the free electrons produced
during the EoR. The corresponding optical depth, $\tau= 0.09 \pm 0.03$ (Spergel {\sl et al.} 2007), 
implies that the intergalactic medium is already significantly reionized at $z=11$.  In the 
next decade, we will learn a lot more by direct observation of the 
redshifted 21-cm emission from the neutral IGM during the EoR. Such instruments as LOFAR, PAST or
MWA will be able to probe the statistical properties of the 21-cm signal, while SKA will be able
to make a full tomography of the IGM up to $z \sim 11$. See Carilli {\sl et al.} (2004) and 
Carilli (2006) for detailed prospects. 

In the last decade, a lot of work has been done, theoretical and numerical, to predict the 
properties of the 21-cm emission and better optimize the design of the future instruments. Madau
{\sl et al.} (1997) and Tozzi {\sl et al.} (2000) present the first theoretical models of 21-cm emission. 
The signal can be seen either in emission or in absorption against the CMB, depending on
the spin temperature of the neutral hydrogen. Interaction with CMB photons couples the spin 
temperature to the CMB temperature in less than $10^5$ years during the EoR, which would make the
 signal undetectable. Fortunately two other processes tend to couple the spin temperature to the
gas kinetic temperature instead. The first is the excitation of the hyperfine transition 
through collisions 
with electrons or other hydrogen atoms (see Kuhlen {\sl et al.} 2006 for numerical simulations), 
which, however, is efficient only in overdense regions (baryonic $\delta\rho/\rho > 5/[(1+z)/20]^2$).
The second is
the pumping of the 21-cm transition by Ly-$\alpha$ photons (Wouthuysen-Field effect, 
Wouthuysen 1952, Field 1959) which requires a threshold value for the local Ly-$\alpha$ flux to be
effective. Consequently, the value of the kinetic temperature of the gas relative to the CMB 
temperature is crucial for determining the 21-cm emission brightness temperature. During the EoR,
the gas is cooling down due to the expansion of the universe faster than the CMB, but is heated by hydrodynamical 
shocks from structure formation, X-ray from pop-III star or quasars and, to a much lesser extend, by 
Ly-$\alpha$ photons (see e.g., Furlanetto {\sl et al.} 2006a). Simple analytical models are now 
available which take into account various source types and formation history (Furlanetto 2006b). 
They usually predict that the signal can be seen in absorption early on, then in emission later 
on. The prediction for the typical amplitude of the differential brightness temperature is a few
$10$ mK. Analytical models cannot, however, take into account the full complexity of 
the 3D inhomogeneous IGM: numerical simulations including the dynamics of structure formation and,
usually as a post-treatment, a full 3D radiative transfer are required.

Dynamical simulations of structure formation have a long history, but cosmological radiative 
transfer simulations are a more recent field of investigation. Simulations showed that the 
ionized
bubbles around the first sources are not spherical: indeed the ionization fronts propagate fast in
the void and more slowly in the high density filaments (see Abel {\sl et al.} 1999 for the
first simulation). The geometry of reionization is now studied in large simulation boxes 
($\sim 100$ Mpc) to get a statistical sample of ionized bubbles of various sizes 
(Iliev {\sl et al.} 2006a, McQuinn {\sl et al.} 2007). However, with such large boxes, even with
very high resolution simulations, small scale density structures, also called minihaloes, are 
not resolved and are often taken into account through a simple clumping factor. The global effect
of minihaloes is to slow down reionization by consuming photons during their photoevaporation.
But a simple, uniform, clumping factor may be insufficient to model the role of minihaloes, and 
more detailed studies have been performed (Ciardi {\sl et al.} 2006, Iliev {\sl et al.} 2005, 
and Shapiro {\sl et al.} 2006 for the impact on the 21-cm signal). Using the density field from
the dynamical simulations and the ionization fraction and gas temperature from the radiative 
transfer simulations, it is possible to produce 21-cm emission maps.

A number of authors provide predictions for different aspects of the 21-cm signal: the emission
map at a given redshift, the average signal as a function of redshift or the signal power as a 
function of the angular scale. These quantities were first predicted from rather small 
simulation boxes ($\sim 10$ Mpc) limiting the angular information to $\theta < 10$ arcmin 
(Ciardi \& Madau 2003, Gnedin \& Shaver 2004, Furlanetto et al. 2004, Valdes et al. 2006).
Recently, predictions from larger simulation boxes ($\sim 100$ Mpc) became available (Mellema 
{\sl et al.} 2006, Iliev {\sl et al.} 2007). It is a fact that the predictions, in particular
the duration and intensity of the absorption phase, depend crucially on the modeling of the 
sources. But other factors have the potential to alter these predictions. Indeed, all predictions
from simulations, at this time, assume a uniform efficiency for the Wouthuysen-Field effect. 
However, Barkana \& Loeb (2005) or Furlanetto {\sl et al.} (2006a) recognized that 
fluctuations in the local Ly-$\alpha$ flux can produce additional fluctuations in the 
brightness temperature. An accurate quantitative modelization of these fluctuations is important. Indeed, the
21-cm signal will be difficult to detect due to brighter foregrounds: the unique signature of its brightness
fluctuations will make the extraction possible. Barkana \& Loeb used simplified analytic models 
(neglecting radiative transfer 
effects on the local Ly-$\alpha$ flux) to compute this contribution. Our goal in this paper is
to investigate how computing the full radiative transfer in the Ly-$\alpha$ line in a 
cosmological inhomogeneous medium can modify the picture presented by these authors.
The 21cm signal will be difficult to detect, due to brighter foregrounds, and the unique character
of its brightness fluctuations is a signature that will make its extraction possible. It is
therefore crucial to precise in more details those fluctuations.

Although Monte-Carlo simulations of Ly-$\alpha$ transfer have a long history, starting with
Avery \& House (1968), only recently has the computing power become sufficient to tackle the
case of a 3D inhomogeneous medium with kinematics, without restrictions on the optical thickness
regime (Ahn {\sl et al.} 2001). Several authors have now developed similar codes to simulate
the Ly-$\alpha$ emission from high redshift galaxies (Zheng \& Miralda-Escud\'e 2002, Cantalupo
{\sl et al.} 2005, Dijkstra {\sl et al.} 2006, Verhamme {\sl et al.} 2006, Tasitsiomi 2006). 

In this paper, we will present the implementation and validation of Ly-$\alpha$ radiative transfer 
in LICORICE, a dynamics-radiative transfer code, with a special emphasis on the treatment large
Hubble flows, which is specific to the EoR applications. 
This work is part of the SKADS\footnote{http://www.skads-eu.org} effort (DS2-T1 task) 
to produce simulated emission maps that
can be used to optimize the design of the future SKA telescope.

In section 2, we present which physics of the Ly-$\alpha$ transfer is included in LICORICE, 
and justify why some aspects of the physics do not need to be included. In section 3 we detail
some aspects of the algorithms and acceleration schemes implemented in the code. Three validation
tests are presented in section 4, comparing the outputs of LICORICE against analytic solutions 
or standard numerical results. Finally, in section 5, we apply the code to some typical EoR 
situation to investigate possible fluctuations in the Wouthuysen-Field effect.

\section{The code: LICORICE } 

LICORICE consists in three parts. A TreeSPH code with multiphase modeling of the gas computes the
dynamics of structure formation (Semelin \& Combes, 2002). A continuum radiative transfer part is
added to compute reionization. This part uses a Monte-Carlo approach and is similar to CRASH (Maselli
{\sl et al.} 2003). It has the advantage over CRASH of using an adaptative grid. LICORICE is 
currently participating in the second part of the Cosmological Radiative Transfer Comparison 
Project (see Iliev et al. 2006b for the first part of the project). The third part is the 
Ly-$\alpha$ radiative transfer which is described in this paper.  

\subsection{Physics of the Ly-$\alpha$ radiative transfer}  

LICORICE implements a Monte-Carlo approach to radiative transfer: it propagates
photons on an adaptative grid. Consequently we will describe the physics of 
Ly-$\alpha$ radiative transfer from the point of view of a photon traveling 
through the simulation box.

\subsubsection{ Computing the optical depth}
A photon propagating through neutral hydrogen has a probability 
$P(\tau)= e^{-\tau}$ of not being scattered after traveling through an optical 
depth $\tau$ from 
its emission point. We consider Ly-$\alpha$ scattering only, so the optical 
depth can be computed as:
\be
 \tau = \!\int_0^l \! \int_{-\infty}^{+\infty} \!\!\!\! ds \, du_{\parallel} \,\, n_{\hbox{\scriptsize HI}} \,\,\,p(u_{\parallel}) \,\, \sigma\left(\nu (1-{v_{\parallel}^{\hbox{\scriptsize macro}}+u_{\parallel} \over c}) \right),  
\ee 
where $\nu$ is the photon frequency in the global rest frame and 
$n_{\hbox{\scriptsize HI}}$ is the 
local number density of neutral hydrogen. $u_{\parallel}$ is the scattering 
atom velocity along the incoming photon's direction in the moving reference frame of the fluid , 
$p(u_{\parallel})$ is the normalized probability distribution for $u_{\parallel}$ ,
$v_{\parallel}^{\hbox{\scriptsize macro}}$ is the gas macroscopic velocity 
along the incoming photon's direction in the global rest frame, and $c$ is the speed of light. Finally 
$\sigma(\nu^{\prime})$ is the Ly-$\alpha$ scattering cross section of a photon 
with frequency $\nu^{\prime}$ {\sl in the atom rest frame}.

The function $p(u_{\parallel})$ usually results from the thermal distribution of the atoms velocity:
\be
p(u_{\parallel}) \,=\, {1 \over \sqrt{\pi} \, v_{th}} \, \exp\left(- { u_{\parallel}^2 \over  v_{th}^2 } \right) \,\,\, \hbox{with} \,\,\, v_{th} =\sqrt{ {2 k_B T \over m_p }}\,\,.
\ee
Some astrophysical systems have significant velocity gradiants on scales below the best possible resolution
of the simulations. In these cases it may be relevant to model the small scale velocity contribution by adding
a turbulent componant to the thermal velocity
dispersion. We did not include any turbulent contribution in this paper.
 
The exact expression of the Ly-$\alpha$ scattering cross section  is given in Peebles (1993). It is 
well approximated by the usual Lorentzian profile:
\be \qquad
\sigma(\nu) \,=\, f_{12} \,{\pi e^2 \over m_e c}\, { \Delta \nu_L / 2 \pi \over (\nu - \nu_0)^2 + (\Delta \nu_L / 2)^2} \,\, ,
\ee
where $f_{12}=0.4162$ is the Ly-$\alpha$ oscillator strength, $\nu_0=2.466 \times 10^{15}$ Hz 
is the line center frequency and $\Delta \nu_L=9.936 \times 10^7$ Hz is the natural line width.

We introduce the dimensionless parameters $x$, the relative frequency shift, and $b$, the natural to Doppler line width ratio:
\be \qquad
x= {\nu-\nu_0 \over \Delta \nu_D} \quad \hbox{with} \quad \Delta \nu_D=  { v_{th} \over c} \nu_0\,\, , 
\ee
\be \qquad
b = {\Delta \nu_L \over 2 \Delta \nu_D } \,\, . 
\ee

Using these notations, we can write the optical depth increment in the gas local rest 
frame ($v_{\parallel}^{\hbox{\scriptsize macro}}=0$) as:
\be \qquad
d \tau  =  ds \, n_{\hbox{\scriptsize HI}} \, {f_{12} \sqrt{\pi} e^2 \over m_e c \Delta \nu_D} H(b,x) \,\, , 
\ee
or in cgs units,
\be \qquad 
d\tau =  ds \, n_{\hbox{\scriptsize HI}} \,\, 6.87 \times 10^{-14} \, \left( { T \over 10^4 } \right)^{-{1 \over 2}} \!\! H(b,x) \,\, ,  
\ee

\noindent
where $H$ is the Voigt function defined as:
\be \qquad
H(b,x) = {b \over \pi} \int_{-\infty}^{+\infty} \!\! {e^{-y^2} \over (x-y)^2 + b ^2} \,dy
\ee 

To compute the Voigt function, we either use the analytic fit given by 
Tasitsiomi (\cite{Tasitsiomi06}), or, in cosmological situations where the Hubble
flow on a scale of the order of the simulation spatial resolution produces a 
frequency shift larger than the Ly-$\alpha$ line width (see section 2.1.3), we use the simple approximation:
\be \qquad
H(x,b) \simeq  \mathrm{max} \left(e^{-x^2}, {b \over \sqrt{\pi} x^2 } \right) .
\label{simpleH}
\ee

\subsubsection{ Effect of the n=2 state splitting, dust and deuterium.}

A Ly-$\alpha$ photon can excite an hydrogen atom from the $1S_{1/2}$ ground 
state to either the $2P_{1/2}$ or the $2P_{3/2}$ state. As discussed by 
Tasitsiomi (\cite{Tasitsiomi06}), the splitting between these two $n=2$ states 
is small: only $10$ GHz, or $1$ km.s$^{-1}$. If the thermal velocity dispersion
 is much larger than $\sim 1$ km.s$^{-1}$ (i.e $T > 100 K$), the level 
splitting is washed out from the radiation spectrum after just one scattering. 
In the case of the Ly-$\alpha$ background radiation during the EoR, the gas 
temperature drops to $\sim 30 K$. However the optical depth as the photon 
redshifts from one side to the other of the Ly-$\alpha$ line is very high: 
$ \sim 8 \times 10^5 $ for the  average gas density of the universe ($\Omega_b=0.045$)
at $z=9.5$ in a standard 
cosmological model. Thus the photon will scatter many times off thermal atoms 
and still the splitting will be washed out. Consequently, we did not 
distinguish between the two $2P$ states in this paper.

Another issue is the possible reshuffling from $2P$ to $2S$ through collisions 
with free protons or electrons. An atom cannot be excited directly to the $2S$ 
state because of the dipole selection rule. But if the $2P \rightarrow 2S $ 
transition is induced by a collision, then the atom will de-excite  through the 
emission of 2 continuum photons: the Ly-$\alpha$ photon is lost. 
Tasitsiomi (\cite{Tasitsiomi06}, see  eq. 26) computes the probability that an atom 
in the $2P$ state will reshuffle to $2S$ before it de-excites normally by 
emitting a Ly-$\alpha$ photon, $p \sim 8.5 \times 10^{-13} n_p ({T \over 10^4})^{-0.17}$, 
where $n_p$ is the proton number density. What can we expect during the EoR ? First, we are interested in the 
Ly-$\alpha$ background in the cold, {\sl neutral} region of the universe, thus $n_p \ll 1$. 
Moreover, even if we assume a non negligible ionization fraction, 
$p \sim 5 \times 10^{-13}$ at $z \sim 10$ for $T=50K$ and for the critical 
density of the universe with 
$\Omega_b=0.045$. This probability must be compared with the average number of 
scatterings a photon undergoes  as it redshifts through the Ly-$\alpha$ line, 
which is of the order of the optical depth: $\sim 10^6$. We see that only a 
fraction of $5 \times 10^{-7}$ of the Ly-$\alpha$ photons will be degraded 
into 2 continuum photons before they redshift out of the line. We conclude 
that this process is not relevant during the EoR.

Dust is usually a factor in Ly-$\alpha$ transfer simulations since it absorbs 
Ly-$\alpha$ photons. Hansen and Oh (\cite{Hansen06}) study the effect of dust 
absorption in a multiphase medium and show that photons can escape such a 
medium, while they would be absorbed in a more homogeneous single-phase medium, 
for the
same total HI column density. In the cosmological context  we are working in 
($z>6$), there are no observations to help us constrain the dust abundance and
 distribution. We can assume that during the EoR, dust may be found only 
around the sources and that the IGM is completely dust free. Under this 
assumption, the effect of dust on the Ly-$\alpha$ flux can be modeled with a 
simple escape fraction coefficient. Furthermore, if this coefficient is assumed to be independent of 
the source, dust will not have any effect on the Ly-$\alpha$ flux {\sl fluctuations} which are the
 focus of this paper. Therefore, we did not include the effect of dust.

Dijkstra {\sl et al.} (\cite{Dijkstra06}) show that the presence of deuterium 
with an abundance [D/H]$= 3 \times 10^{-5}$ leaves a clear imprint on the 
spectrum emerging from a uniform sphere of gas with a central source and a 
total optical depth $\tau = 7.3 \times 10^5$. Is deuterium relevant to the 
Ly-$\alpha$ flux during the EoR? The deuterium line center is $82$ km.s$^{-1}$ 
blueward of the hydrogen line. This is equivalent to the redshift of a photon 
traveling $\sim 0.5$ comoving Mpc at z=9.5. So the first answer is that 
deuterium may have an effect on the Ly-$\alpha$ flux fluctuations only at small
scales ( $< 1$ Mpc comoving). However, let us notice that the total optical depth
for deuterium through the Ly-$\alpha$ line is $\tau \sim 20$ for an abundance
[D/H]$=2 \times 10^{-5}$, and the optical depth in the wing of the Hydrogen Ly-$\alpha$ line, as the 
photon redshifts from far into the blue to the center of the deuterium line 
happens by chance to be also $\tau \sim 20$. This means that, while the photon 
will indeed scatter a few times in the deuterium line, it will also have been
 scattered in the hydrogen line wing several times before it reaches the 
frequency range where deuterium scattering dominates (a
 few 10 km.s$^{-1}$ around the line center). We conclude that the presence of 
deuterium is unlikely to affect the Ly-$\alpha$ flux fluctuations noticeably. 
This does not mean that the effect of deuterium on Ly-$\alpha$ radiation 
cannot be observed during the EoR. Indeed, in the case of an ionizing bubble 
around a source with a sharp ionization front, the continuum spectrum of the 
source will show a Gunn-Peterson trough with a step at the bubble redshift. Deuterium should
 create a secondary small step.

\subsubsection{Dealing with large Hubble flows}

In $z \sim 10$ cosmological simulations, the Ly-$\alpha$ thermal line width is equivalent to 
the Hubble flow redshift over only a few 10 comoving kpc. This scale
 is usually (much) smaller than the size of cells in simulations. So we
must be careful when we compute the optical depth: using a single comoving 
frequency for the photons throughout a cell would result in photons flowing 
through the line core without feeling it. We must actually compute an integral 
along the path of the photon, with a redshifting comoving frequency, to obtain 
the correct optical depth. If we consider that the expansion velocity between 
any two points of the same cell is non-relativistic 
($ {v_H \over c} \sim 0.01$ for 
20 comoving Mpc at $z\sim 10$), the computation gets easier. Let $\nu_{in}$ be the 
comoving frequency and $x_{in}$ the local rest frame value of $x$ when the photon enters the cell.
\be \qquad
x_{in} = {\nu_{in}(1-{\mathrm{\bf v}^{\mathrm{macro}}\cdot \mathrm{\bf k} \over c})  - \nu_0 \over \Delta \nu_D}\,\, , 
\ee
where $\hbox{\bf v}^{\mathrm{macro}}$ is the macroscopic velocity of the gas (uniform inside the cell), 
and  \hbox{\bf k} is the direction of the photon. Then, at a given 
point inside the cell defined by the vector $ r \hbox{\bf k}$ from the entering 
point of the photon, the comoving frequency is:
\be \qquad
\nu = {a(\nu_{in}) \over a(\nu)} \nu_{in} \sim  { \nu_{in} \over 1 + {Hr \over c} } \sim \nu_{in} (1 - {\mathrm{H}r \over c}) \,\, , 
\ee
where $\mathrm{H}$ is the Hubble constant at the simulation redshift, and $a$ is the 
expansion factor. We neglect any variation of $\mathrm{H}$ during the photon travel and 
we consider non-relativistic expansion velocities. 
For non-relativistic macroscopic velocities of the gas, the corresponding value of $x$ writes:
\be \qquad
 x= x_{in}- {\mathrm{H} r \over c} {\nu_{in} \over \Delta \nu_D} \,\, .
\ee

So, we {\bf linearized} the relation between $r$, the 
current path length inside the cell, and the current local rest frame value of 
the $x$ variable. In this approximation, noting $x_{out}$ the value of $x$ when the photon exits 
the cell,  computing the optical depth reduces to computing the integral Voigt function
$H_{\mathrm{int}}(x_{out})$,
with the following definition for the $H_{\mathrm{int}}$ function:

\be \qquad
H_{\mathrm{int}}(x) = \int_{x_{in}}^{x} H(x^{\prime},b) dx^{\prime} \,\, .  
\ee

However, in the Monte-Carlo method, to find the location of a scattering event,
we need to solve {the equation $H_{\mathrm{int}}(x) = A $, where $A$ is a constant}. This is simplified if we
can provide an analytic expression for $H_{\mathrm{int}}(x)$, {with an explicit inverse function}. It is the case if we use 
the simple approximation of $H(x,a)$ given in eq. \ref{simpleH}. It involves the 
$\mathrm{erf}(x)$ function for which we are using an approximation which has an explicit
inverse function.

Modeling the effects of expansion only by a redshift computed from a radial dilatation is the usual 
approximation for Ly-$\alpha$ radiative transfer codes. Other effects of the variation of the expansion
factor $a$, such as variation of the average density during the photon flight time, are usualy ignored.
However, let us emphasize that adding the expansion velocity to the other types of
velocities (macro or microscopic) to compute Doppler shifts, either during 
scattering events or to compute local rest frame values of $x$, works only if 
all velocities are non-relativistic. However, we want to study the
Ly-$\alpha$ flux during the EoR, and at $z \sim 10$, a photon emitted just below
the Ly-$\beta$ frequency will travel $\sim 350$ comoving Mpc and be
redshifted to Ly-$\alpha$ by a Hubble flow
velocity such that $ v_H/c \sim {\delta \nu \over \nu} \sim {\delta a \over a} \sim 0.15$. In this case, the 
second order
errors in computing the redshift are not completely negligible. But, what is more important, neglecting the 
variations of $a$ along the photons path, in computing gas densities for 
example, produces first order error in ${\delta a \over a}$. Consequently, we should limit ourselves to 
$\sim 100$ comoving Mpc boxes.
In this work we do neglect variations of $a$, except for the cosmological redshift, and in most 
cases we use simulation boxes smaller than $\sim 30$ comoving Mpc. The full effect of
expansion will be introduced in the future to handle larger boxes.

\subsection{Scattering off hydrogen atoms}

In an expanding universe the natural variable is the comoving frame frequency 
of the photon. However, as long as the Hubble flow velocities are 
non-relativistic, we can use the frequency in the rest frame of the 
zero-coordinate point of the simulation, hereafter named {\sl global frame},
and treat the expansion as a simple 
radial dilatation with a $ \mathrm{\bf v}_H = H \mathrm{\bf r} $ velocity field to be 
added to the peculiar velocities.  

In the rest frame of the atom, we will consider the scattering to be resonant.
The effect of the recoil, which would change the frequency of the photon by
transferring part of its momentum to the atom, has been shown to be negligible
in astrophysical situations by several authors (Zheng \& Miralda-Escud\'e \cite{Zheng02} or 
Tasitsiomi \cite{Tasitsiomi06}). In the global frame,
however, due to the various contributions to the atoms velocity, the frequency 
$\nu$ of the photon will change. 

There are three main contributions to the atom 
velocity: the thermal motion, the macroscopic peculiar motion and the Hubble 
flow. When the photon scatters off an atom, we first compute the frequency in
the atom rest frame:
\be \qquad
\nu_{\mathrm{atom}}=\nu(1-({\mathrm{\bf v}_H+\mathrm{\bf v}^{\mathrm{macro}}+\mathrm{\bf u}\over c})\cdot \mathrm{\bf k}_i) \,\, ,
\ee
where $\mathrm{\bf u}$ is the thermal velocity of the atom, and 
$\mathrm{\bf k}_i$ 
is the direction of the incoming photon. Let us split $\mathrm{\bf u}$ into
$\mathrm{\bf u}_{||}$, the component parallel to the incoming photon 
direction, and $\mathrm{\bf u}_{\perp}$, the perpendicular component.
 ${ \mathrm{\bf u}_{||} \over v_{th}}$ obeys the distribution:
\be \qquad \label{paralvel}
P_1(y)= {b \over \pi H(b,x)} { e^{-{y^2}} \over (x-y)^2+b^2} \,\, , 
\ee
while each of the 2 components of $\hbox{\bf u}_{\perp} $ obeys:
\be \qquad
 P_2(y)= {1 \over \sqrt{\pi} v_{th}} e^{- { y^2 \over v_{th}^2}}\, \,. 
\ee
Then we compute the direction of 
the photon after scattering. As described by Tasitsiomi (\cite{Tasitsiomi06})
the scattering phase function depends on the excitation state and on whether the 
photon scatters in the wing or in the core of the line. Several authors (see 
e.g.  Cantalupo {\sl et al.} \cite{Cantalupo05} or Verhamme 
{\sl et al.} \cite{ Verhamme06} ) have shown that, for high optical depth media,
the shape of the phase function does not alter the results of the simulation. In this work, 
we use isotropic scattering. Using the new direction of the photon, we then
recompute the frequency in the global frame using a non-relativistic Doppler 
effect.

\section{Ly-$\alpha$ radiative transfer: numerical methods}
\subsection{Gas density and velocity field}
LICORICE is meant to use results from dynamical simulations done using the 
TreeSPH part of the code (Semelin and Combes \cite{Semelin02}). It uses the same Tree structure as the dynamical 
code to build an 
adaptative grid. The grid is build in such a way that each cell contains between
1 and $N_{\mathrm{max}}$ gas particles. Values from 1 to 30 are commonly used for $N_{\mathrm{max}}$.
The density and velocity fields are then interpolated from the particle 
distribution. The cells limit the resolution of the simulation only by having 
uniform dynamical properties. Ly-$\alpha$ transfer inside one cell is still computed 
exactly under the assumption that the expansion velocity on the scale of the cell is non-relativistic.
\subsection{ Monte-Carlo method}
Using the Monte-Carlo approach, we send individual photon from the source and
follow them from scattering to scattering and from grid cell to grid cell, 
until they exit the simulation box. After one event (emission or scattering),
the algorithm is the following:

\begin{itemize}
\item Step 1: Compute the photon global frame frequency, either from the 
scattering atom rest frame frequency, or from the source spectrum.
\item Step 2: Draw the new photon direction (isotropically in this work).
\item Step 3: Draw a variable $p$ from a uniform distribution between $[0,1]$. The photon 
will travel an optical depth $\tau=-\ln(p)$ to the next scattering 
event.
\item Step 4: Increment optical depth with current cell contribution. Determine if scattering occurs
in this cell, if yes go to step 5, if no, pass on to next cell and repeat step 4.
\item Step 5: Draw scattering atom thermal velocity, and compute frequency in the scattering atom rest frame. Go back to step 1.
\end{itemize}
\subsection{Acceleration scheme}

To generate random variable following eq. \ref{paralvel} distribution, we use the method by
Zheng \& Miralda-Escud\'e (\cite{Zheng02}). These authors introduce a parameter
$u_0$ for which an optimal value is needed. We use the following empirical
fit:
\be \quad
$$ u_0=1.85-\log(b)/6.73+\ln(\ln(x)) \quad \hbox{for} \quad x > 3 \, ,$$
\ee
and $u_0=0$ otherwise. This has been determined from a systematic numerical
 optimization, and works well for $10^{-4} < b < 10^{-1}$. For large values of
$x$, quite common in cosmological simulations, drawing a value from
$P_1$ is still slow, even with Zheng \& Miralda-Escud\'e method. In this case,
however, scattering will most likely occur in the wing of the line. 
Consequently, for $x > 10$, we revert to $u_0=0$ but we truncate the 
distribution to the limited range  $[-3,3]$.

We use the core-skipping acceleration scheme (see Avery \& House 1968 or Ahn, 
Lee and Lee 2002 for first applications). In this scheme, we choose a core value
$x_c$ for the variable $x$. In the regime $x < x_c $  the medium must be thick.
If so, the photon will scatter many times over a very short distance. Only when
$x$ gets larger than $x_c$ (scattering in the wing by a fast moving atom) will 
the medium become transparent and will the photon travel a long distance. The 
idea is to ignore the insignificant core scatterings: every time the photon 
enters the core ($x < x_c $), it leaves again immediately by scattering off an
atom with a thermal velocity $u_{at} > x_c v_{th}$. 
{ We use the detailed prescriptions given by Tasitsiomi 
(\cite{Tasitsiomi06}) on choosing $x_c$ as a function
of $b\tau_0$, where $b$ is defined in eq. 5 and $\tau_0$ is the optical depth at the line center . The core-skipping scheme works well when the expected output of the code is
an emerging spectrum. We will show that it also works for computing the fluctuations of the
 local scattering rate. }

\section{Validation tests}
LICORICE being a complex, multipurpose code, we take care of validating each 
part separately. We present here validation tests for the Ly-$\alpha$ part, 
against analytical solutions or standard numerical setups.
\subsection{ Static homogeneous sphere: emerging spectrum}
\begin{figure}[t]
\resizebox{\hsize}{!}{\includegraphics{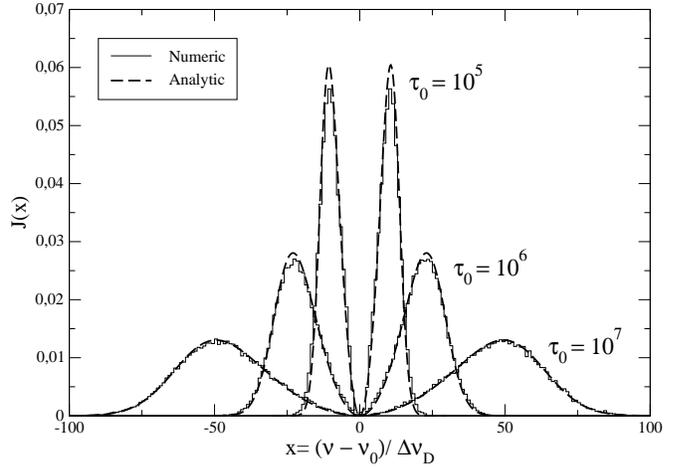}}
\caption{ Emerging spectrum for a uniform and spherical cloud of gas at T$= 10$ K and several 
values of $\tau_0$, the total optical depth at line center.
Numerical histograms are computed with $10^5$ photons. The analytic solution 
is given in Dijkstra {\sl et al.} (\cite{Dijkstra06}).
}
\label{test_Dijkstra}
\end{figure}

A classical validation test for Ly-$\alpha$ codes is the emerging spectrum for
a monochromatic source in the middle plane of a static homogeneous {\sl slab} of
gas. The main reason is that Neufeld (\cite{Neufeld90}) gives an analytic 
solution for the emerging spectrum in the case of an extremely thick system. 
However Dijkstra {\sl et al.} (\cite{Dijkstra06}) provide a new analytic 
expression in the case of a uniform spherical cloud of gas:
\be \qquad
J(x) \,=\, {\sqrt{\pi} \over \sqrt{24} a \tau_0} \left[ { x^2 \over 1+ \cosh\left(\sqrt{{2\pi^3 \over 27}} {|x^3| \over a \tau_0} \right) }\right] 
\ee
Since this geometry is more relevant in a cosmological context, we present
results for this case (we also performed the Neufeld test and also found a 
good agreement). Fig. \ref{test_Dijkstra} shows the comparison between 
numerical and analytic spectra, for a static spherical cloud of gas with 
temperature $T=10K$ and optical depths at line center from center to edge equal
to $10^5$, $10^6$ and $10^7$. For each run, $10^5$ photons are injected at
the center of the cloud with frequency $\nu_0$. { The emerging spectrum $J(x)$ 
is obtained by computing a normalized histogram of the frequencies of the photons
as they leave the cloud.}
The agreement is good in all 
cases, although it gets better and better as $\tau_0$ increases. Indeed in the
case $\tau_0=10^5$, we have $a \tau_0 = 1500$ which is close to the lower limit
for an extremely thick medium.

\subsection{ Expanding homogeneous sphere: emerging spectrum}

\begin{figure}[t]
\resizebox{\hsize}{!}{\includegraphics{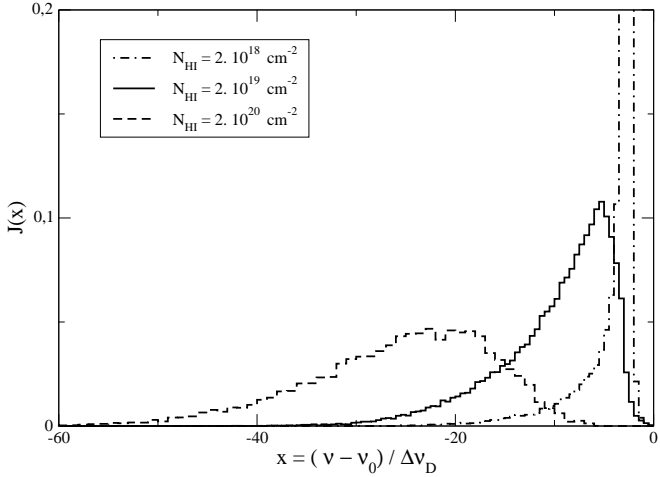}}
\caption{ Emerging spectrum for a spherical, uniform and expanding cloud of 
gas at T$= 20000$ K and several values of $\mathrm{N}_{\mathrm{HI}}$, the column density 
from the center to the edge of the cloud. The emission is at Ly-$\alpha$ frequency from the central
 point of the cloud. The radial velocity of the gas is proportional to the radius and 
equals $200$ km.s$^{-1}$ at the edge of the cloud. The peak 
blueward of the Ly-$\alpha$ frequency is completely suppressed by the expansion. }
\label{test_Z-ME}
\end{figure}

The second test is also becoming a classic. It was first performed by Zheng and Miralda-Escud\'e
(\cite{Zheng02}), followed by several authors. Verhamme {\sl et al.} (\cite{Verhamme06}) present
especially detailed results. The system is a uniform sphere of gas expanding (or contracting)
 with an Hubble-like velocity field: the radial velocity is proportional to the radius. We use
the same physical condition as Zheng and Miralda-Escud\'e (\cite{Zheng02}): a temperature of
$20\,000K$ and a radial velocity of $200$ km.s$^{-1}$ at the edge of the cloud. We use three
different column densities from the center to the edge of the cloud: $N_{\mathrm H}$ equals 
$2. \times 10^{18}$ cm$^{-2}$, $2. \times 10^{19}$ cm$^{-2}$ or $2. \times 10^{20}$ cm$^{-2}$ 
(that is $\tau_0 = 8.3 \times 10^4$, $8.3 \times 10^5$, or $8.3 \times 10^7$). We do not run the tests
for a contracting cloud, and we consider only a central point source emitting at 
Ly-$\alpha$ frequency, not the case of uniform
emissivity. An expanding cloud with a central point source is the most relevant to the EoR. The
 emerging spectra are shown in fig. \ref{test_Z-ME}. We find results very similar to Zheng and 
Miralda-Escud\'e (\cite{Zheng02}) or Verhamme {\sl et al.} (\cite{Verhamme06}). The peak 
blueward of the Ly-$\alpha$ frequency is completely suppressed by the expansion (not shown
in fig. \ref{test_Z-ME}). For the case of a contracting cloud, the red peak would be suppressed

\subsection{ Expanding homogeneous sphere at T=0K: mean intensity field}

Our last validation test relies on an analytic solution given by Loeb and Rybicki (\cite{Loeb99})
. The setup is again an expanding uniform hydrogen medium with a central monochromatic source at 
Ly-$\alpha$ frequency. The velocity field is a Hubble flow again, but this time, the 
gas temperature is $T=0$K. Loeb and Rybicki introduce the dimensionless variable 
$\tilde{\nu} = { \nu_0 - \nu \over \nu_{\star}}$, where $\nu_{\star}$ is the comoving frequency 
shift from Ly-$\alpha$ at which the optical depth to infinity equals 1, and 
$\tilde{r} = {r \over r_{\star}}$, where $r_{\star}$ is the proper radius at which the Doppler
shift from the source due to the Hubble expansion equals $\nu_{\star}$. Loeb and Rybicki give
an analytic expression for the corresponding dimensionless mean intensity $\tilde{J}$, valid in
the diffusion regime:

\be \qquad
\tilde{J}(\tilde{r},\tilde{\nu})\,=\,{1 \over 4\pi} \left( {9 \over 4 \pi \tilde{\nu}^3} \right)^{3 \over 2} \exp \left[ - {9 \tilde{r}^2 \over 4 \tilde{\nu}^3} \right].
\ee

\begin{figure}[t]
\resizebox{\hsize}{!}{\includegraphics{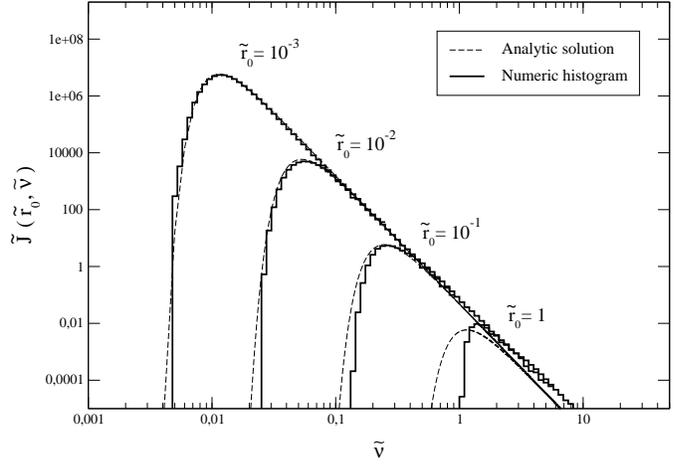}}
\caption{Mean intensity spectra at various radii for a Ly-$\alpha$ monochromatic source in a 
uniform expanding medium at $T=0$K. The numerical result is compared to the analytic solution 
given, in the diffusion regime, by Loeb and Rybicki (1999).}
 \label{test_Loeb_nu}
\end{figure}

\begin{figure}[t]
\resizebox{\hsize}{!}{\includegraphics{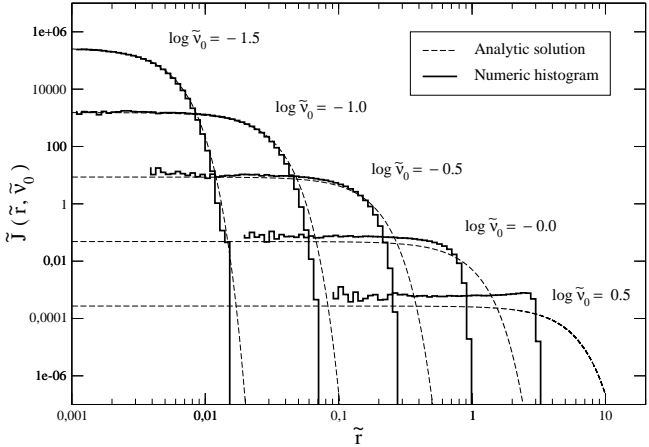}}
\caption{Mean intensity profile at various frequencies for a Ly-$\alpha$ monochromatic source in a
uniform expanding medium at $T=0$K. The numerical result is compared to the analytic solution
given, in the diffusion regime, by Loeb and Rybicki (1999). }
\label{test_Loeb_r}
\end{figure}

Loeb and Rybicki (\cite{Loeb99}) show the comparison between the analytic solution and the 
numerical solution given by a dedicated Monte-Carlo code. Tasitsiomi (\cite{Tasitsiomi06})
runs this test with his general purpose code. Our results are presented in fig. \ref{test_Loeb_nu} 
and \ref{test_Loeb_r}. They are very similar to those of Loeb and Rybicki and Tasitsiomi: 
the numerical results are close to the analytic solution where the diffusion regime is valid. However,
 where photons enter the free streaming regime ($\tilde r_0=1$ in fig. \ref{test_Loeb_nu} and $\log \tilde \nu_0 = 0.5$ in fig. \ref{test_Loeb_r}), the numerical solution diverge from the analytic solution which becomes 
invalid. 

Here are some details on how we ran this test. We used
a temperature $T=2$K for computing the optical depth, which is singular for $T=0$K, but we used a zero thermal velocity for
the atoms in all scattering events. We used a simulation box holding a sphere of gas of radius
$10 r_{\star}$. Since the temperature is not zero in all respects, it reintroduces a dimension
in the problem. Very close to the source the thermal speed is larger than the expansion velocity
and the numerical behaviour should diverge from the analytic solution. For reference, we used a
proper expansion velocity of $200$ km.s$^{-1}$ at $r_{\star}$. In our setup, the thermal
and expansion velocity are of the same order  for $r \sim 10^{-3} r_{\star}$.

The radii range covered by this test is quite large, from $10^{-3} r_{\star}$ to $10 r_{\star}$.
Using $10^{12}$ cells for the radiative transfer grid is not really an option. Actually we used
only $16^3$ cells, but we took advantage of the integral scheme described in equation 12 and 13
for computing exactly the optical depth between two points in the same cell. This approach is 
obviously validated by the good agreement shown in fig \ref{test_Loeb_r} at small radii.

\section{ Ly-$\alpha$ during the EoR}
\subsection{The role of Ly-$\alpha$ photons in the 21 cm emission }
The 21-cm signal can be seen in emission or absorption against the CMB. The differential brightness 
temperature observed at redshift z=0 is:

\be \qquad
\delta T_b = {T_S - T_{\mathrm{CMB}} \over 1+z} (1- e^{-\tau_{21}}) \, ,
\ee
where $T_S$ is the neutral hydrogen spin temperature, $T_{\mathrm{CMB}}$ is the CMB radiation blackbody 
temperature at redshift $z$, and $\tau_{21}$ is the 21-cm line optical depth. The value of 
$\tau_{21}$ is given, among others, by Tozzi {\sl et al.} (2000) or Furlanetto {\sl et al.}
 (2006a). Injecting the formula for $\tau_{21} \ll 1$, the differential brightness temperature 
can be written:

\be \quad 
\delta T_b \sim 9. \,\,x_{\mathrm{HI}}\,\, (1+\delta) (1+z)^{1 \over 2}\,\, {T_S -T_{\mathrm{CMB}} \over T_S} \,\,\,\,{\mathrm{mK}} \,,
\ee
where $\delta$ is the local overdensity at redshift $z$, and $x_{\mathrm{HI}}$ is the neutral 
hydrogen fraction. { This value of $\delta T_b$ is for a flat $\Lambda$CDM model with 
$h_0=0.7$, and $\Omega_b=0.044$. It changes by $\pm \, 0.5$ mK when $\Omega_m$ varies from $0.25$ to $0.3$}. The value of the spin temperature $T_S$ is the result of three competing 
processes. The absorption/reemission  of CMB photons tends to bring $T_S$ to $T_{\mathrm{CMB}}$  over a 
time scale under $10^5$ years during the EoR (Tozzi {\sl et al.}, 2000).  As we have seen
in the Introduction section, both collisions between hydrogen atoms 
and the pumping by Ly-$\alpha$ photons, also known as the Wouthuysen-Field effect 
(Woutuysen 1952, Field 1958), tends, instead, to couple $T_S$ to the kinetic temperature of the 
gas. As a result, the spin temperature can be written (Furlanetto {\sl et al.} 2006a):

$$
T_S^{-1}={ T_{\mathrm{CMB}}^{-1} +x_c T_K^{-1} + x_{\alpha} T_C^{-1} \over 1+x_c+x_{\alpha}} \quad \mathrm{with} \quad T_C \simeq T_K \,, $$
where $x_c$ and $x_{\alpha}$ are the coupling coefficients respectively for collisions and 
Ly-$\alpha$ pumping, {and $T_C$ is the effective color temperature of the UV radiation field (see Furlanetto {\sl et al.} 2006a)}. The coefficient $x_{\alpha}$, which is the focus of this work, can be 
explicitly written as:

$$
x_{\alpha}={4 P_{\alpha} T_{\star} \over 27 A_{10} T_{\mathrm{CMB}} } \,\, ,
$$
where $T_{\star}=0.068$K, $A_{10}= 2.85 \times 10^{-15} s^{-1}$ is the spontaneous emission 	
factor of the 21-cm transition, and $P_{\alpha}$ is the number of scatterings of Ly-$\alpha$ 
photons per atom per second. Now usually come two approximations that we will {\bf not} make in this 
paper. {\bf The first} is that $P_{\alpha}$ is considered proportional to $J(\nu_\alpha)$,
the angle averaged specific intensity at the local Ly-$\alpha$ frequency, neglecting the 
contribution of wing absorptions. We will see that this approximation is valid provided $J(\nu)$
itself has been computed taking into account wing absorptions. {\bf The second}, more drastic 
approximation, is to evaluate $J(\nu)$ without performing the full radiative transfer 
computation. Actually, to our knowledge, all numerical simulations of 21-cm emission consider a
uniform value of $J(\nu_\alpha)$. However, Barkana and Loeb (2005) have shown that several 
factors induce fluctuations in $J(\nu_\alpha)$: the $1/r^2$ scaling of the flux which magnifies
the Poisson noise in the source distribution, the clustering of the sources, and the 
contribution of higher Ly-$\alpha$ series photons (also studied in detail by Pritchard and 
Furlanetto 2006). They predict the power spectrum of the 21-cm brightness temperature due to the 
fluctuation in $J(\nu_\alpha)$. Although a vast improvement over using a uniform Ly-$\alpha$ flux, 
they are still neglecting radiative transfer effects: they assume that photons are freely 
streaming until they redshift to the local Ly-$\alpha$ frequency. We will show that this 
assumption  breaks down at scales smaller that $\sim 10$ comoving Mpc.

Potentially, pumping the upper excitation level is not the only way in which Ly-$\alpha$ 
photons can influence the 21-cm emission: they also heat up the gas.  This heating mechanism was
first thought to be efficient by Madau {\sl et al.} (1997). However, Chen \& Miralda-Escud\'e (2004), 
taking into account the atoms thermal velocity distribution which had been neglected by
Madau {\sl et al.}, found a much smaller, actually negligible heating rate. Furlanetto \& 
Pritchard (2006c), taking into account the effect of higher Lyman series photons confirmed this
result and found, for typical EoR conditions, the Ly-$\alpha$ heating rate to be 140 smaller than
the heating rate from X-rays.
Chuzhoy \& Shapiro (2007a) recently challenged this result, adding in particular the 
effect of the deuterium $Ly-\beta$ resonance line, but the strength of this effect
is not yet completely probed.

\subsection{$P_\alpha$ profiles for spherically symmetric configurations}

In the next three cases, we consider a central source in a spherically symmetric medium of neutral 
hydrogen at $T_K=30$K, and $z \simeq 10$, which is typical of the EoR . The main difference with 
the setup of the tests in section 4.3 is that the source emits a continuous flat spectrum. We 
deal only with photons between Ly-$\alpha$ and Ly-$\beta$ frequencies.

\subsubsection{Homogeneous medium}

\begin{figure}[t] 
\resizebox{\hsize}{!}{\includegraphics{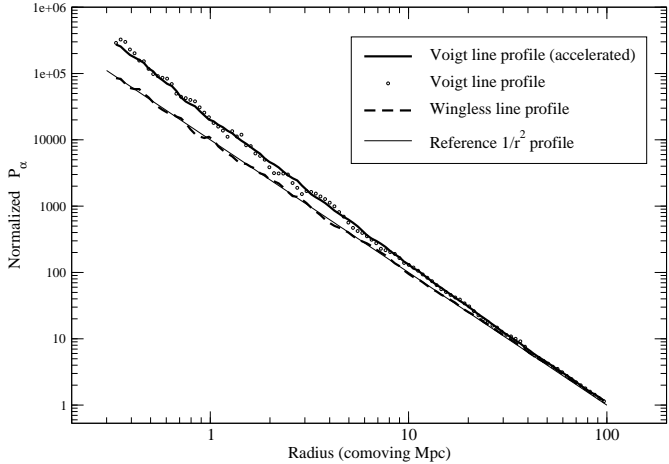}}
\caption{ Profiles of the scattering rate per atom,$P_\alpha$, in the Ly-$\alpha$ line, around a
central source with a continuous flat spectrum in an homogeneous medium of neutral hydrogen at 
30K and the average baryon density of the universe at $z\sim10$.}
\label{radial_homog}
\end{figure}  
First, we consider the case where the gas is homogeneous with a density equal to the average 
density of the universe  at $z \sim 10$. We ran the simulation in a 600 comoving Mpc box with the source in the 
center of the box. We present the results up to a radius of 100 comoving Mpc: at larger radii 
the effect of the box boundaries (no retrodiffusion) and the variation of $a$ during the photon flight 
(see section 2.1.3) alter the validity of the simulation. Fig. \ref{radial_homog} presents the 
radial profile of the scattering rate per atom, $P_{\alpha}$, in different cases. Using the real
Voigt line profile, we can see a deviation at small scale from the simple $1/r^2$ profile
 expected if photons stream freely from the source until they redshift to the local Ly-$\alpha$ 
frequency. Here is why. Let us consider a photon emitted  above the Ly-$\alpha$ frequency. As it 
travels away from the source, it is redshifted toward the local Ly-$\alpha$ frequency. Because of
the contribution of the wings of the Voigt profile to the optical depth, it has a probability to 
scatter before reaching the local Ly-$\alpha$ frequency. On average, it will scatter for the first time, 
when the optical depth along its path reaches 1. This is achieved when the photon is 
redshifted to the frequency $\sim \nu_\alpha+\nu_\star$ (see section 4.3), which occurs at a distance
$r_\star$ before the location where it would reach the local Ly-$\alpha$ frequency 
(quantities defined in Loeb \& Rybicki 1999). In our setup, typical of the EoR, 
$r_\star \sim 10$ comoving Mpc. After the photons scatter for the first time they change 
direction. Consequently, the location where a photon actually enters the core of the local 
Ly-$\alpha$ line has a probability to be anywhere within $\sim 10$ comoving Mpc of the location 
determined by free streaming alone. This is not crucial at large scales where the expected 
$1/r^2$ profile is recovered, but it creates a steeper profile at small scales 
(exponent $\sim -2.3$ in our setup). To validate our interpretation, we computed the transfer 
with a modified line profile: the core of the line is unchanged, but the wings are set to zero.
As can be seen in Fig. \ref{radial_homog}, we then recover the $1/r^2$ profile.
{After this paper was submitted, Chuzhoy and Zheng (2007b) posted a paper with a similar 
result. They considered a very similar setup and computed the transfer with a simple 
Monte Carlo code which is 1-D (spherical symmetry) and uses a simplified line profile but does 
include Ly-$\alpha$ photons locally injected by cascades from upper Lyman series lines. They find
the same steepening of the scattering rate profile at short scales. However, they show that 
photons injected from upper Lyman series lines are much less sensitive to radiative transfer 
effect: they follow the $1/r^2$ profile more closely. Consequently the discrepancy between
the full radiative transfer computation and the simple $1/r^2$ evaluation is somewhat reduced.}

Finally we checked the effect of using the core-skipping acceleration scheme for evaluating the
fluctuations of $P_\alpha$: since all core absorptions are avoided, it could have 
modified the spatial $P_\alpha$ fluctuation map. As can be seen in Fig. \ref{radial_homog}, it 
is not the case. Although the wings modify the location where photons enter the core, the scattering 
number is still dominated by the close-to-the-core contribution.

\subsubsection{Central clump}

\begin{figure}[t]
\resizebox{\hsize}{!}{\includegraphics{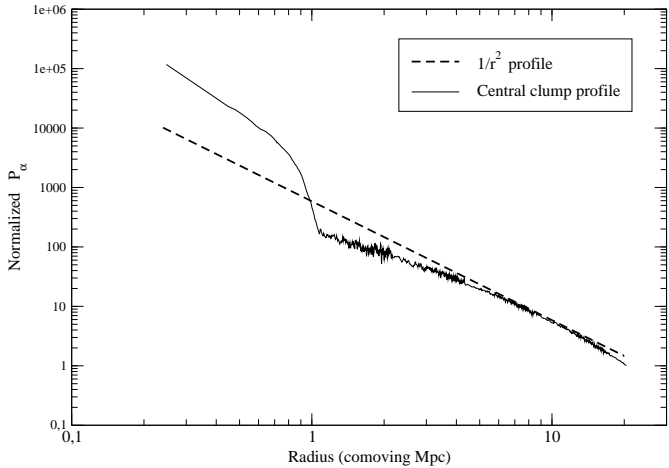}}
\caption{Profile of the scattering rate per atom, $P_\alpha$, in the Ly-$\alpha$ line, around a
central source with a continuous flat spectrum, inside an overdense spherical clump of gas (64 
times the density of the surrounding medium) of radius 1 comoving Mpc,  }
\label{central_clump}
\end{figure}  
Until now, the Ly-$\alpha$ scattering rate has been evaluated during the EoR in an homogeneous 
medium only (Barkana and Loeb 2006, Pritchard and Furlanetto 2006). With a general purpose 3D 
code such as LICORICE, we can investigate the impact of density fluctuations in the gas on the
Ly-$\alpha$ scattering rate {\bf per baryon}. First we choose a very simple setup: we consider a
box of size 64 comoving Mpc, with a central homogeneous spherical clump of gas of radius 
1 comoving Mpc. The clump is 64 times denser than the surrounding medium, which has the average baryon
density of the universe at the redshift of the simulation ($z\sim10$). The source is in the center of the clump and has 
a continuous flat spectrum. The radial profile of the Ly-$\alpha$
scattering rate per atom, $P_{\alpha}$, is shown in Fig. \ref{central_clump}. The main feature is
a depletion of $P_{\alpha}$ in the low density medium just outside the clump. Indeed, photons
that should redshift to the Ly-$\alpha$ frequency in this region, have first to travel through the
high density region where they have an enhanced probability to be scattered in the wing of an 
atom, and redirected to redshift to local Ly-$\alpha$ while still inside the core. In other 
words, the enhanced wing scatterings in the clump draw core scatterings  from surrounding 
regions to the clump itself. This is somewhat similar to the usual shadowing effect in 
radiative transfer, although in this case the process occur in the frequency space so the {\sl shadowing} 
can be seen even in a spherical configuration. Obviously, in addition to the $1/r^2$ 
decline of the flux, and the fluctuations in the source distribution, we can expect a new source
of fluctuations for $P_\alpha$: the density fluctuations of the intergalactic medium.

\subsubsection{Isothermal density profile}

\begin{figure}[t]
\resizebox{\hsize}{!}{\includegraphics{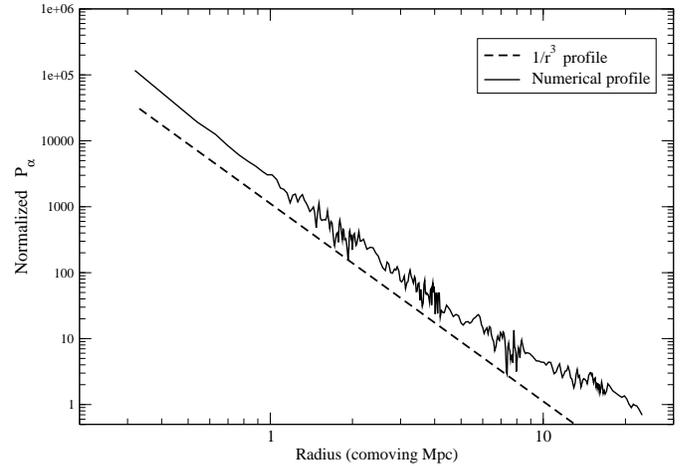}}
\caption{Profile of the scattering rate per atom $P_\alpha$ in the Ly-$\alpha$ line, around a
central source with a continuous flat spectrum. Inside a sphere of radius 10 comoving Mpc, the
gas density field is an isothermal sphere, outside it is homogeneous at the average baryon 
density of the universe ($z\sim 10$). The two regions connect smoothly. A $1/r^3$ profile is plotted for comparison.}
\label{isothermal_profile}
\end{figure}

Instead of the sharp density transition between the clump and the surrounding medium we now 
consider an isothermal density profile ($\rho \sim 1/r^2$) around the source. The isothermal
profile connects to the surrounding homogeneous medium at a radius of 10 comoving Mpc. All 
other parameters of the simulation are the same as in the previous setup. The radial profile of 
the Ly-$\alpha$ scattering rate per atom, $P_{\alpha}$, is shown in Fig. \ref{isothermal_profile}.
Since the average density is higher than in the other setups, photons scatter many times and even
with a acceleration method, we used only $2 \times 10^4$ photons. To avoid a very high noise level 
 at large radii we used an adaptative resolution: high in the center, lower in the outer 
regions. We can see on the figure that, inside the region with an isothermal density profile, 
$P_\alpha$ closely matches a $1/r^3$ profile. It reverts to $1/r^2$ in the homogeneous medium 
outside the $10$ Mpc radius. This shows that the brightness temperature fluctuations of the 
21-cm emission may be stronger than previously estimated at small scales, at least during the
early EoR when $P_\alpha$ fluctuations are meaningful.

Is this setup more relevant to the prediction of $P_{\alpha}$ 
fluctuations during the EoR than a uniform medium ? Yes in the sense that the medium should 
obviously be denser closer to the source. However the spherical symmetry and the specific 
profile used here are oversimplified: the actual intergalactic medium in the EoR is not in an 
equilibrium configuration, especially during this early period of sources formation, and the 
spherical symmetry is broken by the presence of filaments. In the next section we investigate how
filaments modify $P_{\alpha}$.

\subsection{$P_\alpha$ map for an axisymmetric configuration}

\begin{figure}[t]
\resizebox{\hsize}{!}{\includegraphics{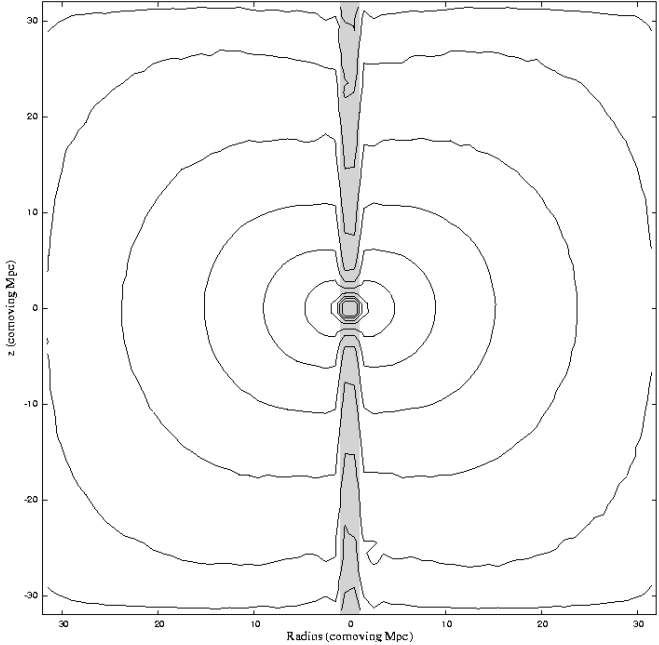}}
\caption{Contours of the scattering rate per atom $P_\alpha$ in the Ly-$\alpha$ line, around a
central source with a continuous flat spectrum. The source is located inside a  cylindrical 
homogeneous filament of gas (shaded zone in the figure) with a density 64 times the density of 
the surrounding medium which is at the average baryon density of the universe ($z \sim 10.$).
The contours are equally spaced on a logarithmic scale with a step of $\sqrt{10}$. }
\label{contour}
\end{figure}

\begin{figure}[t]
\resizebox{\hsize}{!}{\includegraphics{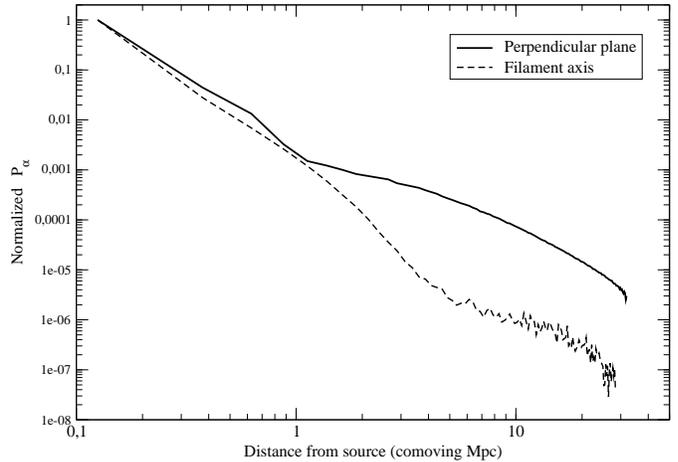}}
\caption{Scattering rate per atom $P_\alpha$ in the Ly-$\alpha$ line along the axis of a filament and in a 
perpendicular plane. A central source with a continuous flat spectrum is located at the intersection of the 
filament and the plane. The filament 
is cylindrical with radius 1 comoving Mpc and homogeneous with a density 64 times the density of
the surrounding medium which is at the average baryon density of the universe ($z \sim 10.$).}
\label{filament_profiles}
\end{figure}

We consider now an axisymmetric density field for the gas: inside a cylinder of radius 1 
comoving Mpc the density is 64 times the critical baryon density, outside it is equal to the
critical baryon density. The source is located on the symmetry axis. All other parameters are 
identical to the previous setup. This setup is a simplified representation of the real case of
source formation during the EoR at the intersection of several filaments,
where the filaments have a density profile. The $P_\alpha$ contour map is shown in Fig. 
\ref{contour}. The shaded area represents the filament, and the contours are equally spaced on a
logarithmic scale (two per decade). The map is integrated over a variation of $\pi$ of the 
angular variable. About $5. 10^6$ photons where used for this simulation. There is some boundary
effect due to the finite size of the simulation box: photons that would scatter just outside the simulation 
box and possibly reenter the box are lost instead. This affects a few Mpc near the boundary.
We can see a sharp
depletion in the number of scatterings per atom inside the filament. We have checked that a
smaller density contrast creates a smaller depletion, as expected. Once again, we see that
the gas density field fluctuations induce fluctuations in $P_\alpha$. However, the weaker
Wouthuysen-Field effect in denser regions may be balanced by the greater coupling due to 
collisions (which is proportional to the density). What may be more relevant to the future 
observations than what occurs at small scales inside the filaments, is that the presence of the 
filaments modifies the shape of the contours in the low density surrounding medium: they are not
spherical but oblate. With our density contrast, the axis ratio is about 2 in the 5-10 Mpc range
and decreases at large distances.
Fig. \ref{filament_profiles} shows the normalized $P_\alpha$ profiles along filament axis and in the 
perpendicular plane containing the source. While inside the filament (distance smaller than 1 comoving Mpc), 
the profiles match. At larger distances, fig. \ref{filament_profiles} quantify the $P_\alpha$ ratio between
the two regions: it reaches a maximum value of $\sim 100$ at a distance of $\sim 5 Mpc$.

\section{ Conclusions}

The main goal of this paper was to investigate a source of fluctuations in the brightness 
temperature of the 21-cm emission during the EoR usually neglected in numerical simulations. The 
Wouthuysen-Field effect, which couples the hydrogen spin temperature to the kinetic temperature
of the gas, is regulated by the crucial parameter $P_\alpha$: the number of scatterings per atom
per second. While all previous simulations of the 21-cm emission used a uniform value for 
$P_\alpha$, Loeb and Barkana (2005) have shown, in a simple theoretical framework, that several 
sources of fluctuations in $P_\alpha$ can modify the power spectrum  of the 21-cm emission. We 
studied how a full 3D radiative transfer treatment of the Ly-$\alpha$ line in a cosmological 
context modifies the picture given by Loeb and Barkana.

The first step was to implement and validate the Ly-$\alpha$ radiative transfer in LICORICE. We
used a Monte-Carlo approach, and implemented an algorithm and acceleration schemes similar to
those of other existing codes. We discarded physical processes such as recoil or deuterium 
contribution, which have negligible effect for the Ly-$\alpha$ transfer during the EoR. On the 
other hand we took care to compute the optical depth accurately in an expanding cosmological
medium, without
any resolution effects. We presented three validation tests for the code. The first two are 
classical setups: a monochromatic source in a static uniform sphere of gas, and in an 
expanding sphere of gas. The agreement for the emerging spectrum with analytic solutions and 
results by other authors is good. The third test, the mean intensity map for a monochromatic 
source in an expanding sphere of gas at $T=0$K, focuses on a quantity more closely related to 
$P_\alpha$. The comparison with the analytic solution provided by Loeb and Rybicki is good where
the analytic solution is valid: in the diffusion regime. This set of tests strongly suggests 
that the LICORICE is valid.

Barkana and Loeb (2005) show that several factors contribute to the fluctuations in the local
Ly-$\alpha$ flux: the $1/r^2$ scaling of the flux, and the Poisson noise and clustering in the 
sources distribution. They assumed, however that photons redshift freely from the source until 
they reach the local Ly-$\alpha$ frequency, and only then scatter off hydrogen atoms. In other
words, they neglected wing scattering. We computed the $P_\alpha$ profile for a source with a flat
continuous spectrum in a uniform expanding medium. We showed that the effect of taking into 
account the wing scatterings in a full radiative transfer code is to steepen the profile at small
scales to a $\sim 1/r^{2.3}$ profile (with our choice of parameters). At large scales, or when
wings are suppressed, we recover the $1/r^2$ profile. Thus we may expect, at small scales, more
power in the 21-cm emission than predicted by Barkana and Loeb. But a yet stronger effect was
obtained when we introduced fluctuations in the density of the gas surrounding the source. We
investigated a central clump  setup and an isothermal profile. In both cases we observed 
alterations in the $P_\alpha$ profile. In the case of an isothermal density profile we found 
a $\sim 1/r^3$ profile for $P_\alpha$. This also suggests stronger fluctuations than predicted
by Barkana and Loeb, but, once again, mainly at small scales since the $\sigma_8$ at $z=10$ is
only $\sim 0.1$: large scale structures had not  enough time to grow yet.

Finally, we tried  to create a more realistic situation by placing a flat spectrum source inside
a filament of overdense gas. We observed a sharp depletion of $P_\alpha$ inside the filament. 
The photons scatter out of the filament before they reach the core of the line. This 
fluctuation inside the filament is once again a small scale feature, and may be difficult to
catch with LOFAR or SKA. However, the presence of the filament also produced oblate contours
for $P_\alpha$ at larger scales ($> 10$ comoving Mpc), in the surrounding medium. This effect
may be more within reach of the resolution of these instruments.

{The increased fluctuations of $P_{\alpha}$ due to radiative transfer effects and to the
inhomogeneous distribution of the gas translate linearly into fluctuations of the $x_\alpha$
coefficient. On one hand, we expect these fluctuations to be globally significant and produce 
brightness temperature fluctuations only as long as
$x_\alpha \sim 1$, i. e. as long as the coupling does not saturate to $T_s = T_K$. This
occurs in the early phase of reionization. It is not possible to be much more specific in terms of
redshift because the Ly-$\alpha$ pumping efficiency depends strongly on the model for the source 
type and formation history. On the other hand we showed that density fluctuations in the
gas can create fluctuations of $P_{\alpha}$ of a factor greater than $10$. The depleted regions
will fill up only when the amount of young sources gets more than $10$ times larger. Cosmological
simulations suggest that this corrresponds to a change of redshift between 1 and 2 around redshift 10. Consequently we expect that the fluctuations of $P_{\alpha}$ due to inhomogeneous gas will
lead to a longer survival of depleted regions where $T_S$ remains coupled to $T_{\mathrm{CMB}}$.  One may argue that the strongest fluctuations of the gas density are located around the sources
and are ionized very early. However numerical simulations have shown that the ionization front
is not spherical: it is trapped in the high density regions such as filaments pointing to the 
source, where reionization is much delayed (see for example, Gnedin 2000). So, these filaments should be able to play their role
in creating $P_\alpha$ fluctuations.  }

We have not investigated the effect on $P_\alpha$ of an anisotropic peculiar velocity field 
around the source. In principle, it would also induce non spherical contours. However, cosmological 
simulations in a $20$ Mpc box suggest peculiar velocities of the order of $~100$ km.s$^{-1}$ 
during the EoR, when the Hubble constant is H(z=10) $\sim 1000$ km.s$^{-1}$.Mpc$^{-1}$. Moreover, 
only velocity
differences will alter $P_\alpha$ contours, which should be smaller than $~100$ km.s$^{-1}$ at large scales.
We estimate that the impact of the velocity field of the gas is smaller than the impact of the
density fluctuations. However, LICORICE fully implements the effects of the gas peculiar velocity,
and it will be taken into account in the future simulations of a cosmological box.

Another process will have to be included in the future: the effect of higher Lyman series lines.
Barkana \& Loeb (2005) and Pritchard and Furlanetto (2006) have shown that these lines, having
horizons closer to the source than the Ly-$\alpha$ line, add to the power of the $P_\alpha$ 
fluctuations close to the sources. In a forthcoming paper, we will apply LICORICE to a 
cosmological field during the EoR using simulation outputs from the HORIZON 
project\footnote{http://www.projet-horizon.fr}, and compute the resulting 21-cm brightness temperature map. If,
as we believe, we find significant modification of the predictions, we will include 
higher Lyman series lines.

\begin{acknowledgements} 
This work was realized  in the context of the SKADS and HORIZON projects. 
\end{acknowledgements}

\end{document}